\documentclass[a4paper,10pt]{article}
\usepackage{graphicx}
\usepackage{siunitx}
\usepackage{hyperref}
\usepackage[english]{babel}
\usepackage[latin1]{inputenc}
\begin{document}
\title{Constraining chemical networks in Astrochemistry}

\author{S. Viti \footnote{viti@strw.leidenuniv.nl}, J. Holdship\\
Leiden Observatory, Leiden University, PO Box 9513, 2300 RA Leiden, \\The Netherlands\\
    Department of Physics and Astronomy, University College London,\\ Gower Street, London, WC1E 6BT}
\date{}

\maketitle

\section*{Abstract}
Databases of gas and surface chemical reactions are a key tool for scientists working in a wide range of physical sciences.  In Astrochemistry, databases of chemical reactions are used as inputs to chemical models to determine the abundances of the interstellar medium. Gas chemistry and, in particular, grain surface chemistry and its treatment in gas-grain chemical models are however areas of large uncertainty. Many reactions - especially on the dust grains - have not been systematically experimentally studied. Moreover, experimental measurements are often not easily translated to the rate equation approach most commonly used in astrochemical modelling. Reducing the degree of uncertainty intrinsic in these databases is therefore a prime problem, but has so far been approached mainly by ad hoc procedures of essentially trial and error. In this chapter we review the problem of the determination of accurate and complete chemical networks in the wider context of Astrochemistry and explore the possibility of using statistical methods and machine learning (ML) techniques to reduce the uncertainty in chemical networks.
\section{Introduction}
The space between the stars, the interstellar medium (ISM), is far from empty. The ISM is populated with interstellar clouds and filaments, made of gas ($\sim$99\%) and dust ($\sim$1\%). These clouds and filaments are the birth sites of stars \cite{1,2}.  Modelling the physical and chemical processes leading to the formation of stars and, hence, planets is a highly non-linear, very time-dependent, problem, involving a multi-layered interconnection between the physics and the chemistry of the gas and the dust. Despite the many advances driven by state-of-the-art telescopes, we are still far from solving the `star formation problem': stars form in molecular filaments or clumps, cold ($\sim$ 10 K) and relatively dense (\textgreater \num{e4} particles \si{\per\centi\metre\cubed}) regions of the interstellar medium, where most of the gas is molecular. These regions also contain higher density ones, called cores. The first two stages of the formation of solar-like systems involve highly non-linear physical and chemical processes that are hard to corroborate by astronomical observations alone. In the first stage, some cores can become gravitationally unstable and initiate star formation. This collapse is controlled by the presence of molecules, which cool the gas and allow gravity to overcome hydrostatic equilibrium. During this pre-stellar core phase, due to the high density and low temperatures in the cores, species from the gas phase `freeze' onto the dust grains present, and form an icy mantle. The gas and surface compositions during these two stages exhibit a complicated time dependent, non-linear chemistry that strongly depends on the physical environment\cite{3}. Little experimental information is available for the interstellar ices: what is the unprocessed ice composition? What are the efficiencies of the viable surface reactions? And how do the energetics of the ISM (cosmic rays, UV radiation, shocks) influence the processed ices?\par
Examples of key issues involve even the simplest molecules: the CO molecule sticks efficiently to surfaces at temperatures below $\sim$ 25 K and is abundant in the ices. Some of this CO can be converted to other species, and the observation of CO$_2$ and CH$_3$OH in ices\cite{4} suggests that this processing does occur. H$_2$CO, CH$_3$OH and glycolaldehyde are involved in the surface hydrogenation of CO, and determining their chemistry in star-forming regions is vital for studies of prebiotic chemistry\cite{5}. Some ices return to the gas phase when the gas temperature rises above 20 K. At lower temperatures, non-thermal desorption processes can still return molecules from solid to gas-phase\cite{6}. However, these non-thermal mechanisms `compete' with the freeze-out. Gas-phase chemistry is better constrained than surface chemistry; nevertheless, many uncertainties still remain on the formation and efficiencies of routes for some of the most complex species (routinely called Complex Organic Molecules, or COMs, by astronomers) observed in space\cite{7,8}.  In summary, the composition of the gas and icy mantles varies according to a time-dependent process highly dependent on the conditions of the gas and dust in any particular cloud.\par
In the second stage, i.e. when the protostar is born, the gravitational energy is converted into radiation and the envelope around the central object, the future star, warms up. The molecules frozen on the grain mantles during the previous phase acquire mobility and likely form new, more complex species. When the temperature reaches the mantle sublimation temperature of $\sim$100 K, the molecules in the mantles are injected into the gas, where they react and form new, more complex, molecules. Simultaneously, a fraction of matter is violently ejected outward in the form of highly supersonic collimated jets and molecular outflows. When the outflowing material encounters the quiescent gas of molecular cloud, it creates shocks, where the grain mantles are (partially) sputtered and the refractory grains are shattered. Once in the gas phase, molecules can be and are observed via their rotational lines. Again, as during the first stage, the interaction of gas and dust, and hence the gas composition, varies within very short timescales (less than one hundred years) and the effects of chemistry and dynamics are interlocked in a complex non-linear fashion\cite{9}.\par
In both stages, molecules provide an essential tool for the analysis of the chemical and physical conditions of star and planet forming regions. Each stellar or planetary evolutionary stage is characterized by a chemical composition, which, if properly interpreted, leads to the determination of the physical processes of its phase.\par
This brief description of the multi-million years star formation cycle highlights our challenge: the interconnection and non-linear correlation of the many parameters with each other or with extra, unknown ones makes determining and specifying the parameter network within chemical models a highly challenging task.\par
Of particular importance is the reliability of the chemical reactions datasets. Databases of chemical reactions and rate coefficients for both gas and solid phase reactions are the key input to all chemical models. While there are some gas phase databases online, in reality each research group creates a personal version of such databases, especially as there is no accepted standard for databases; in addition, there is no database containing a complete set of solid phase reactions. We explore the completeness and reliability of such databases in the next section.\par
\section{Completeness and reliability of chemical reaction databases}
Typically, chemical models of the ISM in use today employ databases of mainly 2-body (sometimes 3-body) chemical reactions; these contain long lists of gas and solid phase chemical reactions. In the gas-phase, for each reaction three constants are provided that are then used to calculate the rate coefficient. The first constant (usually $\alpha$) represents the rate at 300 K and the other two constants ($\beta$ and $\gamma$) give the temperature scaling of that rate. For photo-reactions, an equivalent set of constants gives the rate in an unshielded interstellar UV field and the extinction dependence of that rate. These three constants will have their own associated uncertainty. Depending on the physical conditions, between 20 and 50\% of the reactions may not have been experimentally studied. In some cases, chemical reaction rates are therefore highly speculative.\par
For reactions occurring on the dust grains, the pathways, efficiencies and branching ratios are even more uncertain than for the gas phase reactions, due to the lack of experimental data. Surface reactions are therefore highly speculative and consequently are often not included in astrochemical models.\par
Reducing the degree of uncertainty for all the reactions would require an unfeasibly large number of laboratory experiments. Yet, chemical models fully rely on these databases, making their accuracy a prime problem for Astrochemistry. In the next two sections we will describe in a bit more detail the two categories of databases: gas-phase and surface reaction networks.
\subsection{Gas-Phase Networks}
All the astrochemical models developed through the years calculate the abundances of hundreds of species involved in thousands of chemical reactions. Gas-phase reactions are the backbone of chemical networks as they are the primary routes for the formation and destructions of most molecules. Hence a poor understanding of the rate coefficients can lead to large errors in the abundances of the main molecules observed in the ISM.\par
The range of kinetic temperatures and gas densities over which we need accurate rate coefficients will depend on the region of space we need to model but can be from 10 K to \textgreater 1000 K and from 100 to \SI{e4}{\per\centi\metre\cubed} respectively. Chemical reactions, under these conditions, can be studied in the laboratory or theoretically. However, each experiment, or study, takes at least a few months to complete. Often then rate coefficients are estimated from other already known rate coefficients. More importantly, even if every reaction were to be experimentally or theoretically investigated, it would be impossible to do so under the whole range applicable to the ISM and many extrapolations are therefore performed. Measures have been taken by the community to compile databases where some quality assurance is performed. For example, the Kinetic Database for Astrochemistry (KIDA)\cite{10} is a public gas-phase network which provides all the rate coefficients from various sources and evaluates their accuracy where possible. We refer the reader to \cite{10} for a detailed explanation of this network. However, even the best efforts at transparency and completeness such as KIDA fall short of providing the user with a complete, accurate and reliable gas phase network that includes all the possible species involved in the complex chemistry routinely observed in the ISM. This is especially true for complex organic molecules, which are species that contain 6 or more atoms, are present in the ISM mostly with low abundances and yet are key for our understanding of pre-biotic chemistry in space\cite{7}.  Gas-phase networks including the formation and destruction of COMs are far from complete\cite{11,12} and astrochemical models have been struggling to explain the observed abundances of COMs. While most current models tend to favour grain surface over gas-phase chemistry in COMs formation, several studies have now shown that COMs may indeed form in the gas phase but that, due to the incompleteness of gas phase networks involving COMs, the relative contribution of gas-phase versus surface reactions cannot be quantified.
\subsection{Grain-Surface Networks}
There is no doubt that dust grain chemistry plays a pivotal role in the formation of key abundant species observed routinely in the ISM e.g. H$_2$O, CH$_3$OH and NH$_3$ which are primarily formed through solid state chemistry\cite{13,14}. Although a debate exists on whether gas phase reactions contribute to the formation of COMs, it is clear that the latter are at least partly formed on the surface of the grains.\cite{7,8,15}
We know that dust grains act as `catalysts' meaning that - on their surface - surface reactions and energetic radiation can synthesize molecules as complex as prebiotic species\cite{16,17} starting from very simple molecules (e.g. CO and N$_2$) and atoms (H, C, O, N, S etc) deposited from the gas when the temperature of the medium is close to 10 K. In the last couple of decades many experiments have been performed to evaluate surface chemical reactions (see review \cite{18}). 
Molecular hydrogen was the first molecule to be studied on dust surfaces\cite{19}. Many experiments since then have been performed to study the formation of more complex molecules (e.g. \cite{20, 21, 22}) as well as the ice morphology and ice mantle mechanisms (e.g. \cite{23,24}). However, all these experiments are performed within a constrained range of laboratory conditions which differ from those found in the ISM: for example,  atomic fluxes, ice temperatures, ice morphologies, and mixture ratios, energetic processes in the laboratory will differ from those found in the ISM. Hence formation, desorption and destruction routes and rates for surface molecules as derived from experiments may not always be exhaustive to the needs of the chemical modellers. In other words, experimental data for interstellar ices are limited, since the experimentation process is neither simple nor fast.  As a consequence, most chemical models either include very simple surface reaction networks or ones where most of the reactions are essentially guesswork.
\section{A Bayesian Approach}
In astrochemistry, it is not common to use Bayesian methods as a means of deriving posterior probability distributions (PPDs) for model parameters from observations. The first study to do so \cite{9} used Bayesian inference to derive parameters such as the gas density and cosmic-ray ionization rates within a dark molecular cloud from observations of species in the gas and ices using chemical models. This study particularly highlighted the inverse nature of the astrochemical problem at hand, namely that in typical astrochemical problems we have to deal with nonlinear ill-posed inverse problems where solutions may not be unique or may not depend continuously on the observational data. In one study\cite{9}, a Bayesian approach based on the use of the Metropolis-Hastings (MH) algorithm (an example of a Markov chain Monte Carlo algorithm) was adopted and they used a simplified version of UCLCHEM (a time-dependent gas-grain chemical model, now open source\footnote{\url{https://uclchem.github.io}}) to explore a nine-dimensional parameter space for molecular clouds. They ran two identical sets of eight MCMC chains differing only in the way the prior distribution information was set, with the first set having a non-informative prior information in the form of acceptable range of possible values for observed ices, while in the second the prior include all the observational constraints from observations, including gas phase abundances. Of particular relevance to this review, it is interesting to see how the different prior distributions have affected the resulting Posterior Probability Distributions (PPD) for one of their dimensions: the branching ratios for some key chemical reactions pathways e.g. those that control how much of the oxygen that freezes on the grains turns into H$_2$O or OH. With noninformative uniform prior, the high-density regions of the PPD cover excessively large sections of the distribution, meaning that the branching ratio could not be constrained enough. With informative priors, the high-density region in the PPD is reduced and gives a clear indication that the production of water over OH is favoured. This example showed for the first time in Astrochemistry that branching ratio parameters can be successfully estimated through Bayesian MCMC methods. 
 
A more recent work\cite{25} used the above methodologies as a proof of concept in an attempt to infer rates of reactions for a limited surface network, and in doing so provided the means to derive reduced networks in the context of observational constraints. In this work, a simple chemical model was developed that considers only the solid state chemistry in the ice mantles of dust grains in a dark molecular cloud. The simplified model is a time-dependent single-point model that generates a time series of solid phase molecular abundances as a function of the physical conditions of the molecular cloud and the chemical parameters of the defined chemical network. This network only included surface reactions and mainly hydrogenation of common gas phase species. We report these reactions in Table~\ref{table:reacs}. 

\begin{table}[t]
\centering
\begin{tabular}{c c c c c c}
\hline\noalign{\smallskip}
No. & \multicolumn{5}{c}{Reactions}\\
\hline\noalign{\smallskip}
1.&O &$+$  & H  &$\rightarrow$& OH \\ 
2.&OH &$+$  & H  &$\rightarrow$& H$_2$O \\
3.&CO &$+$  & OH  &$\rightarrow$& CO$_2$ \\
4.&S &$+$  & H  &$\rightarrow$& HS \\
5.&HS &$+$  & H  &$\rightarrow$& H$_2$S \\
6.&H$_2$S &$+$  & S  &$\rightarrow$& H$_2$$S_2$ \\
7.&CS &$+$  & H  &$\rightarrow$& HCS \\
8.&HCS &$+$  & H  &$\rightarrow$& H$_2$CS \\
9.&CO &$+$  & S &$\rightarrow$& OCS \\
10.&OCS &$+$  & H  &$\rightarrow$& HOCS \\
11.&H$_2$S &$+$  & CO &$\rightarrow$& OCS \\
12.&H$_2$S &$+$  & H$_2$S  &$\rightarrow$& H$_2$$S_2$ \\
13.&H$_2$$S_2$ &$+$  & CO &$\rightarrow$& CS2 $+$ O \\
14.&H$_2$S &$+$  & O  &$\rightarrow$& SO$_2$ \\
15.&C$S_2$ &$+$  & O  &$\rightarrow$& OCS $+$ S \\
16.&CO &$+$  & HS  &$\rightarrow$& OCS \\
17.&S &$+$  & O  &$\rightarrow$& SO \\
18.&SO &$+$  & O  &$\rightarrow$& SO$_2$ \\
19.&SO &$+$  & H  &$\rightarrow$& HSO \\
20.&HSO &$+$  & H  &$\rightarrow$& SO \\
21.&CO &$+$  & H  &$\rightarrow$& HCO \\
22.&HCO &$+$  & H  &$\rightarrow$& H$_2$CO \\
23.&H$_2$CO &$+$  & H  &$\rightarrow$& H$_3$CO \\
24.&H$_3$CO &$+$  & H  &$\rightarrow$& CH$_3$OH \\
\hline
\end{tabular}
\caption{he reactions included in the study by \cite{25}. All the reactions are occurring on the grain surfaces.}
\label{table:reacs}
\end{table}
As one can see from this Table only reactions among four types of atoms are considered: O, C, S and H. Obviously the reactions included are not exhaustive of all the possible combinations: the criteria used to choose which reactions to include were based on (i) simple hydrogenation until saturation e.g. reactions 4 and 5 and (ii) reactions that have been found to be efficient, or even dominant, routes to forming a species e.g. reactions 21-24 to form methanol. For this toy model, gas phase reactions were ignored but the depletion of the gas phase on the surface was parameterized.
 
\begin{figure}
    \centering
    \includegraphics[width=\textwidth]{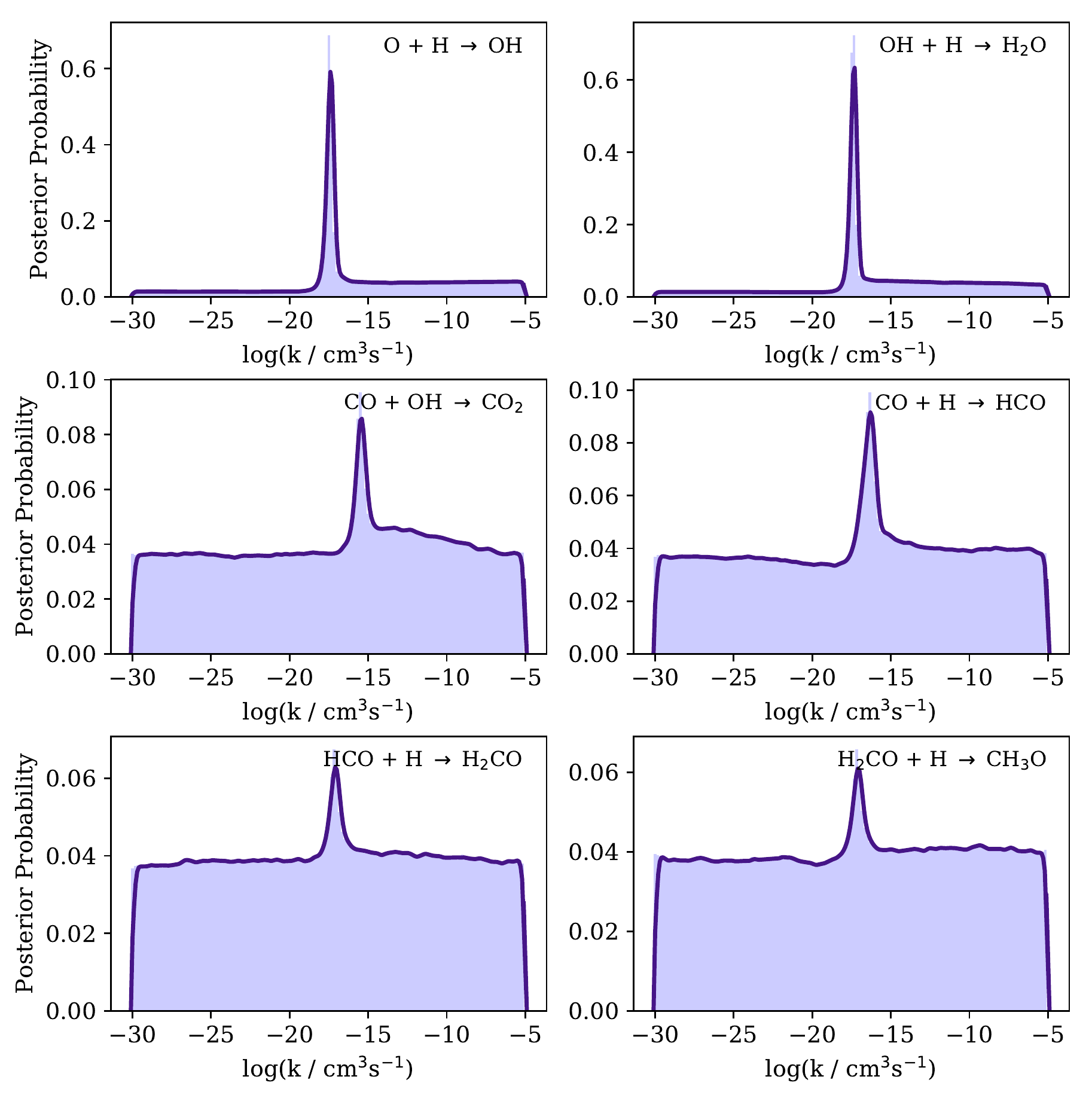}
    \caption{The probability distribution of selected reaction rates, shown as marginalized posteriors.}
    \label{fig:posteriors}
\end{figure}

The result of this inference was the probability distribution of the reaction rates, shown as marginalized posteriors in Fig~\ref{fig:posteriors}. 
They find that when observational constraints directly constrain the abundance of a species involved in a reaction, the range of likely values for the rate of that reaction can be sufficiently narrowed for use in chemical models. Therefore, this method represents a method by which reliable reaction rates can be obtained for use in modelling until laboratory measurements or theoretical work supersede them.\par
Whilst this approach has many strengths, it requires a large amount of computational power. In both of the above examples, the chemical model used for the inference was greatly simplified in comparison to a typical model used to interpret observations. Where the model itself is not the subject of the inference (as it is in the work of \cite{25}), we might use machine learning to map from model inputs to outputs in order to do more efficient parameter inference on the input parameters.

\section{Machine Learning Techniques for complex networks}
Ultimately chemical modellers need to be able to rely on complete as well as accurate chemical reaction networks. Reducing the degree of uncertainty for such networks requires cutting edge Artificial Intelligence techniques combining Artificial Neural Networks, Bayesian Inference methods and advanced Monte Carlo sampling algorithms.  While we are still quite a way from achieving these, ML techniques have now started being used in Astrochemistry.
For example, one work \cite{26} used a neural network to emulate a full chemical model. By creating a training set from the outputs of a full model (UCLCHEM \footnote{\url{https://uclchem.github.io}}) for a wide range of input physical parameters, they were able to train a neural network to predict these outputs. This allows for the computation of chemical abundances required to interpret observations in milliseconds rather than the minutes of CPU time required to run a full chemical model. Consequently, computationally intensive parameter inference can be performed using a close approximation of a full chemical model without taking an infeasible amount of CPU time.
 
Such neural networks would have uses beyond direct interpretation of observations. Many radiative hydrodynamical simulations use highly approximate radiative-chemical treatments to provide the cooling rates for the simulated gas. This is a necessity when a full chemical model would add to the computational burden of an already intensive model. However, evaluating a neural network is generally a small task and as such, more accurate chemistry could be inserted into physical models in the form of pre-trained neural network emulators.
 
The use of machine learning techniques on chemical networks could be of particular importance for the reduction and accuracy of networks of complex organic molecules, in particular those that are as yet undetected in the ISM. For example, the `holy grail' for astrochemists is to find amino acids in space.  Amino acids are important organic compounds that play a key role in the formation of proteins. In the protein production process, amino acids join together to form polymer chains, which represent the structural units of proteins. It is believed that the formation of amino acids may indeed have occurred in the ISM, since they have been firmly detected in meteorites\cite{27}. This is supported by laboratory experiments where amino acids are found to form on the surface of model dust grains by UV-photon and ion irradiation of suitable precursor molecules under astrophysically relevant conditions \cite{28,29}.  Despite great efforts towards their detection, a firm detection of amino acids in the ISM is still - unfortunately - eluding us. 
One possible solution to this is to use ML tools to identify weak signals from the amino acids in spectroscopic telescope data. When detecting molecules, strong peaks in the spectra from an observed object are identified by matching their frequency to known emitting frequencies of molecules. If the emission is weak it may not be detected by a human, or the statistical significance of many weak lines may be missed. A classification model trained to categorize spectra as containing a particular molecule or not, based on the spectral noise and the emission profile at various frequencies, could give confident detections in cases that a human would miss. In the case of previously undetected molecules, such as amino acids, a training set can be produced by creating many synthetic spectra. Further, such an analysis could be performed on thousands of targets with little effort which has made it an interesting prospect for the exo-planetary community aiming to identify molecules in planetary atmospheres\cite{30}.\par
Another solution is to focus our search by determining the optimal conditions for observing amino acids. It is likely that, if present, such amino acids must form on the mantle of dust grains. It is therefore essential we determine the optimal conditions under which simple amino acids, such as glycine and alanine, can form on the dust icy mantles and can subsequently be released into the gas phase.
 
Key prerequisites for understanding how and where glycine and alanine form are the determination of the surface reactions that lead to their formation as a function of (i) kinetic temperature; (ii) gas density, (iii) structure of the gas, (iv) presence and abundance of other molecular species on the ices, (v) UV and cosmic ray ionization rates, among other factors. To date, no large-scale predictions of the formation routes of glycine and alanine have been made, as a function of the parameters above.\par
A possible method to achieve this may be based on advanced and cutting edge techniques developed in the information sciences, ML and statistical disciplines. It should be possible to devise and apply a combination of statistical and ML techniques to perform large scale chemical models, involving large datasets of gas- and surface-phase chemical reactions to derive the physical and chemical conditions under which every surface chemical reaction is viable. A technique worth investigating is probabilistic graphical modelling, a branch of ML that studies how to use probability distributions to describe a particular problem whose `model' has many uncertainties. \par
Ultimately - for any well defined species set (COMs, amino acids, ice species etc.) we need to be able to simultaneously investigate the paths and efficiencies of their formation and destruction over a large physical (densities, temperatures, radiation fields, and cosmic ray ionization rates) and chemical (rate coefficients) parameter space, and discover if, where and under what conditions these species are abundant in space.
\section{Conclusion}
Machine Learning and probabilistic methods for solving typical astrochemical problems is a fast-growing field. As larger chemical reaction networks and more complex models are being employed in astrochemistry, the need for intelligent data mining algorithms will increase. It is clear that there are large uncertainties, as well as lack of information on formation and destruction routes, in gas and especially surface reactions and rate coefficients. To date, these uncertainties have only been tackled by laboratory experiments and quantum mechanical calculations. However, the complexity of the reaction networks, the lack of much prior information in the form of observed ices, and the length of each laboratory experiment, make a parameter exploration (in terms of gas densities, gas temperatures, UV and cosmic ray fluxes for example) unfeasible. Initial `experimentation' with ML algorithms is proving to be an efficient avenue to tackle this challenge.

\end{document}